\documentclass[12pt,preprint]{aastex}

\shorttitle{Two Clusters With Radio Quiet Cooling Cores}
\shortauthors{Donahue et al.}

\usepackage{graphics,graphicx,epsf,natbib}
\received{}

\def\Del{\Delta}

\begin{document}
\title{Two Clusters of Galaxies with Radio-Quiet Cooling Cores}
\author{Megan Donahue and G. Mark Voit}
\affil{Michigan State University, Physics \& Astronomy Dept., East Lansing, MI 48824-2320}
\email{donahue@pa.msu.edu, voit@pa.msu.edu}
\author{Christopher P. O'Dea}
\affil{Dept. of Physics, Rochester Institute of Technology,
84 Lomb Memorial Dr., Rochester, NY 14623-5603}
\author{Stefi A. Baum}
\affil{Center for Imaging Science, Rochester Institute of Technology,
54 Lomb Memorial Dr., Rochester, NY 14623-5603}
\author{William B. Sparks}
\affil{Space Telescope Science Institute, 4700 San Martin Dr, Baltimore, MD 21212}

\begin{abstract}
Radio lobes inflated by active galactic
nuclei at the centers of clusters are a promising candidate for
halting condensation in clusters with short central cooling times because they
are common in such clusters.  In order to test the AGN-heating hypothesis, we obtained 
{\em Chandra} observations of two clusters with 
short central cooling times yet no evidence for AGN activity:  Abell 1650 and Abell 2244.
The cores of these clusters indeed appear systematically different from cores with
more prominent radio emission.  They do not have significant central temperature gradients,
and their central entropy levels are markedly higher than in clusters with stronger radio
emission, corresponding to central cooling times  $\sim 1$~Gyr.  Also, there is no
evidence for fossil X-ray cavities produced by
an earlier episode of AGN heating.  We suggest that either 
(1) the central gas has not yet cooled to the point at which feedback 
is necessary to prevent it from condensing, possibly because it is conductively stabilized, 
or (2) the gas experienced a major heating event $\gtrsim 1$~Gyr in the past and has 
not required feedback since then.  
The fact that these clusters with no evident feedback have higher central entropy 
and therefore longer central cooling times than clusters with obvious AGN feedback 
strongly suggests that AGNs supply the feedback necessary to suppress condensation
in clusters with short central cooling times. 
\end{abstract}

\section{Introduction}

The cooling-flow problem in clusters of galaxies has been one of the most
notorious issues in galaxy formation.  
The cooling time ($t_c$) of gas within the central $100-200$~kpc of many 
clusters is less than a Hubble time 
\citep[e.g. ][]{CowieBinney77, FabianNulsen77}.  
If there is no compensating heat source distributing 
thermal energy over that same region, that gas ought to cool, condense, and relax 
toward the cluster's center in a so-called  ``cooling flow,''  
but exhaustive searches in other wave bands have failed
to locate the $10^{12}-10^{13} \, M_\odot$ of stars or cool gas that should have 
accumulated 
\citep[e.g. ][]{ODea1994, Antonucci1994, McNamaraJaffe1994}. 
Nevertheless,  something unusual is happening in clusters with $t_c \ll H_0^{-1}$. 
Significantly smaller amounts of gas have been detected in the form of 
CO \citep{Edge2002, EdgeFrayer2003} or HI \citep{1994ApJ...436..669O},
vibrationally excited H$_2$ \citep{Donahue2000, Falcke1998, JaffeBremer97},  
and evidence for star formation 
\citep[e.g. ][]{Cardiel1998,Crawford1999, VD97, ODea2004}
are common in these systems, and {\em Chandra} observations have 
shown that radio lobes sometimes carve out huge cavities in
the X-ray emitting gas at the centers of such clusters 
\citep[e.g., ][]{McNamaraA2597_2001, Fabian2000_NGC1275, Blanton2003}. 
  
This association of star formation, line emission, and 
relativistic plasma with cooling-flow clusters has fed 
speculation that feedback from active galactic nuclei 
modulates the condensation of hot gas, greatly 
reducing the mass-cooling rates naively inferred from 
X-ray imaging \citep[e.g., ][]{Bohringer2002, Quilis2001}.  
However, active feedback sources 
are not found in every cluster with $t_c \ll H_0^{-1}$. 
For example, the nearby cooling-flow sample of 
\citet{Peres1998}
 consists of twenty-three clusters with 
$\dot{M} > 100 \, M_\odot \, {\rm yr}$ inferred from {\em ROSAT}
imaging.  Of these, thirteen have both an emission-line nebula 
and a strong radio source, two have no emission lines but a strong 
radio source (A2029, A3112), and three have emission lines but a 
weak radio source (A478, A496, A2142) leaving five with no 
emission lines and little or no radio activity (A1651, A2244, 
A1650, A1689, A644).

To test the idea that feedback from either an AGN, star formation,
or some combination of the two suppresses cooling in the cores of 
clusters with $t_c \ll H_0^{-1}$, we observed two objects from
this last set of five with {\em Chandra}:  A1650 ($z=0.0845$) and A2244 ($z=0.0968$). 
These clusters are luminous X-ray 
sources, with bolometric  
$L_x \sim 8 \times 10^{44} h_{70}^{-2}$ erg s$^{-1}$ and
estimated gas $T_x$ of 5.5-7.0 keV \citep{David1993}.
Here we compare those clusters with an archival sample of clusters of
similar X-ray luminosities ($L_x=0.4-30 \times 10^{44}$ erg s$^{-1}~h_{70}^{-2}$)  and temperatures
($T_x=2.9-7.4$ keV), with 
$t_c \ll H_0^{-1}$,  with evidence for active feedback in the form of central 
radio emission, and in most cases, with emission-line nebulae as well 
 \citep{Donahue2005B}.  We will refer to these clusters as "active clusters." 
All of the clusters in the \citet{Donahue2005B} sample and the two clusters
discussed in this paper have single, optically luminous, brightest central galaxies 
residing at the centroid of their X-ray emission.
\S~2 describes the observations and calibration procedures.
\S~3 describes the data analysis and the extraction of entropy 
profiles, and \S~4 discusses our results, which we summarize
in \S~5. For this paper we assume $H_0=70$ km s$^{-1}$ Mpc$^{-1}$
and a flat universe where $\Omega_M=0.3$.

\section{Observations and Calibration}

The observation dates, flare-free exposure times, and count rates between 0.5-9.5 keV within
a 4' radius aperture are reported in 
Table~\ref{log}. The back-illuminated CCD on the Chandra X-ray Observatory
\citep{Chandra2002}, the ACIS-S3 detector, was used for its sensitivity
to soft X-rays. Its field of view ($8\arcmin \times 8\arcmin$) 
extends to about 10\% of the virial radius of each cluster, limiting our analysis
to the cluster cores.

\begin{deluxetable}{lccc}
\tablecaption{Chandra Observations \label{log}}
\tablehead{
\colhead{Cluster} & \colhead{Observation Date} & \colhead{Exposure Time} &
\colhead{ACIS-S Count Rate} \\ 
 \colhead{}             &        \colhead{}                             &  \colhead{(s)}   & \colhead{(ct s$^{-1}$)} }
\startdata
Abell 1650 &  Aug 3-4, 2003      &  27,260        &     4.36                \\
Abell 2244 &  Oct 10-11, 2003   &  56,965        &     4.35               \\ 
\enddata
\end{deluxetable}

We processed these datasets using the {\em Chandra} calibration software 
CALDB 2.29 and CIAO 3.1, released in July 2004\footnote{Chandra Interactive
Analysis of Observations (CIAO), http://cxc.harvard.edu/ciao/}. 
Neither observation experienced flares.
We used Chandra deep background observations for our background spectra.\footnote{M. Markevitch, author of http://asc.harvard.edu/cal/ Acis/Cal\_prods/bkgrnd/acisbg/ COOKBOOK}
Source and background spectra were extracted using identical concentric 
annuli containing a minimum of 20,000 counts per source spectrum. 
Bright point sources were excluded from the event files
before spectral extraction. The spectra were binned to a minimum of 25 counts per 
energy bin. 

Using XSPEC v11.3.1, we fit the projected and deprojected spectra from 0.7-7.0 keV to  
MekaL models \citep{MekaL} with Galactic absorption attenuating the
soft X-rays \citep{MM1983}. 
Since the best-fit absorption overlapped the Galactic values of $N_{\rm H}=1.56$ and $2.3 \times 10^{20}$
for A1650 and A2244 respectively \citep{DickeyLockman1990}, we fixed $N_{\rm H}$ at
those values for this analysis. The positions of the Fe-K lines were consistent with the cluster redshifts from galaxy velocities in \citet{StrubleRood1999}.
We computed 90\% uncertainties ($\Del \chi^2 =2.71$) for the temperature, 
normalization, and metallicity at each 
annulus. We constrained the metallicity to be 
constant across 2-3 annuli.  The reduced $\chi^2$ values for the fits
were typically 1.10-1.15. More details about our data analysis strategy and further analyses 
are described in \citet{Donahue2005B}, where
we also analyze Chandra archival observations of nine other cooling-flow clusters that have central
radio sources and emission-line nebulae.

Neither cluster exhibits a strong temperature gradient across the core. Abell 2244
is nearly isothermal with $kT = 5.5 \pm 0.5$ keV at every radius $<4'$, and Abell 1650 
varies from $5.5 \pm 0.5$ keV in the core to $7.0 \pm 1.0$ keV at 4', statistically 
consistent with but somewhat higher than the temperature profile over similar radii obtained with 
XMM measurements.  
(\citet{TakahashiYamashita2004} adopted a lower redshift ($z=0.0801$) to fit the XMM data, 
 which may indicate a calibration uncertainty.) These small inner temperature gradients contrast with those of most other cooling flow clusters, 
which tend to be more pronounced.   Both cluster cores contain a significant metallicity gradient, 
ranging from 0.6-0.8 solar in the center to a more typical 0.2-0.3 solar outside the core, 
consistent with \citet{TakahashiYamashita2004}. This metallicity 
pattern is typical of the other cooling flow clusters we studied.

\section{Data Analysis}

The goal of our data analysis was to determine whether  or not clusters without
obvious signatures of feedback were systematically different  from those with radio sources
that do show signatures of feedback.  Our primary results are that these two clusters
do not show any evidence for ghost cavities and have higher central entropy levels
than clusters showing evidence for feedback.

In order to search for cavities in the intracluster medium, we adaptively smoothed
the X-ray data to a minimum significance of 5-sigma with both a Gaussian and a top-hat kernel.
We found no ``ghost bubbles."  On scales larger than about 50 kpc from the center, both clusters exhibited regular, nearly round intensity contours. We also did not see evidence for 
filaments, such as that found tracing the H$\alpha$ emission in 
Abell 1795 \citep{2001MNRAS.321L..33F} or M~87 \citep{Sparks2004}.

We determined the entropy profiles of these clusters by computing the 
adiabatic constant $K = kTn_e^{-2/3}$ at each radius to quantify the 
specific entropy.  The temperature ($kT$) profiles were measured as described 
in \S2.  The electron density profiles ($n_e$) were derived by deprojecting 
the 0.5-2.0 keV surface brightness profiles within 
annuli having 5" widths using the technique of \citet{KCC1983}. The 
uncertainties of the deprojected count rate profiles were estimated by bootstrapping 
1000 monte-carlo simulations of the original surface brightness profiles. 
A spatially-dependent  conversion of 0.5-2.0 KeV count rates to electron 
densities was obtained from the X-ray spectroscopy.  For this paper,
we assumed that the temperature and the count-rate conversion factor in the
central bin were constants.

\begin{figure}
\includegraphics*[width=0.5\textwidth,angle=0]{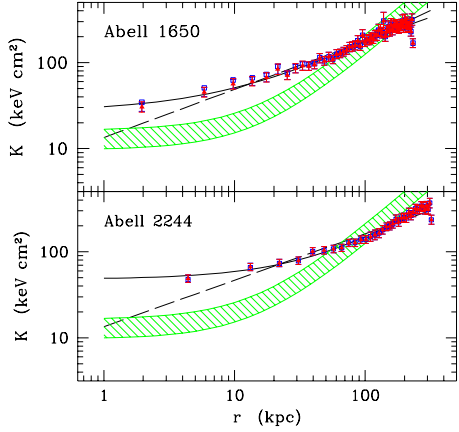}
\caption{Entropy profiles for Abell 1650 and Abell 2244. All gas at $\sim5$ keV with entropy $\lesssim170$ keV cm$^2$ has $t_c < H_0^{-1}$.  The hatched region shows the locus of 
entropy profiles for active clusters from the sample of \cite{Donahue2005B}.
 \label{EntropyProfiles}}
\end{figure}

\begin{figure}
\includegraphics*[width=0.5\textwidth,angle=0]{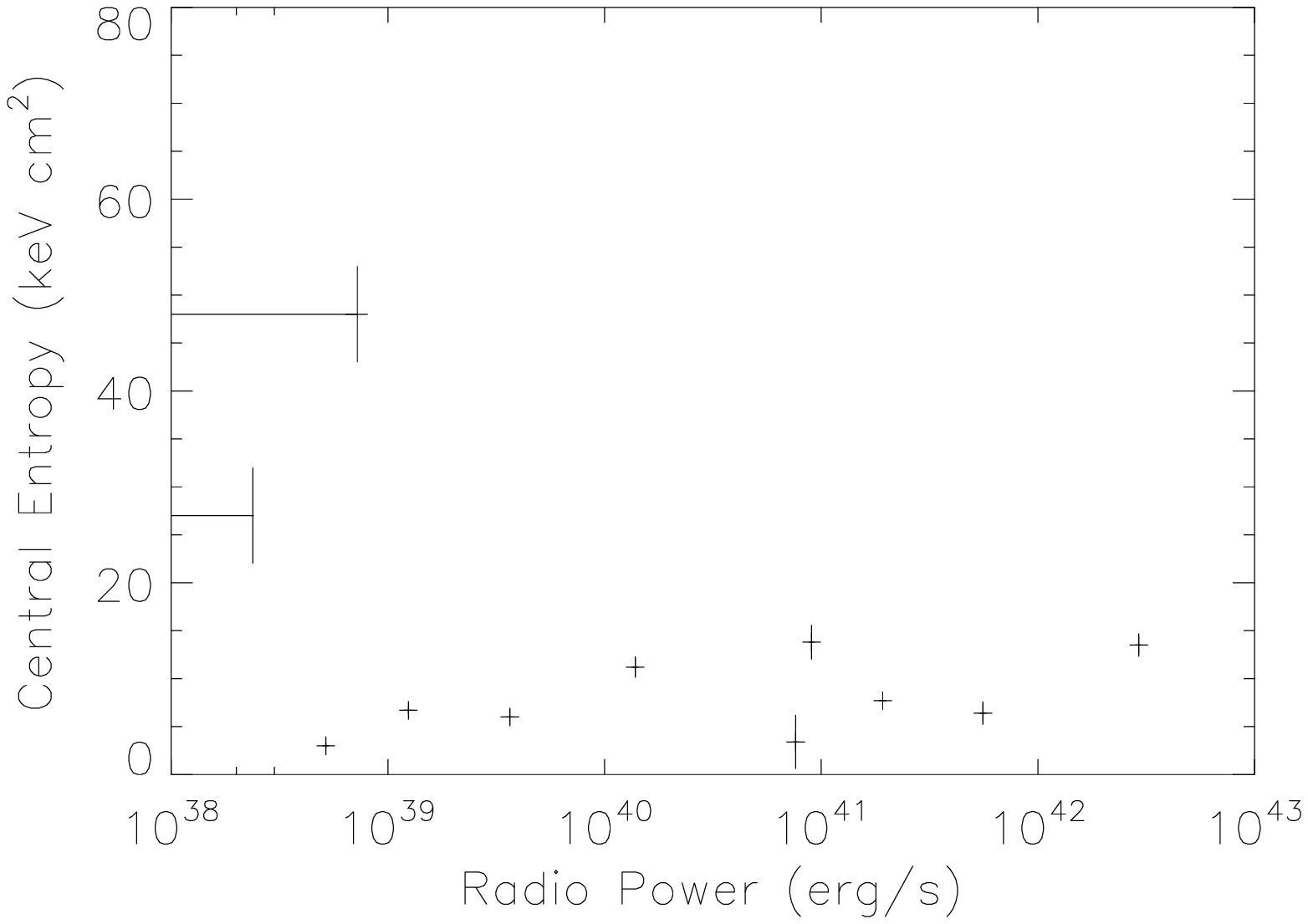}
\caption{ Radio power $\nu L_\nu$ from 20 cm observations \citep{NVSS}
(See also \citet{LedlowOwen1995} and \citet{SBO1995}, and 6 
cm upper limits for A2244 and A1650 from \citet{Burns1990}.) 
\label{figure:EntPower}  }
\end{figure}

Figure~\ref{EntropyProfiles} shows that the entropy profiles of Abell 1650 and Abell 2244 are systematically different from the nine cooling-flow clusters in the sample of active clusters from \cite{Donahue2005B}.
The two radio-quiet clusters have flatter entropy profiles with larger values of central entropy.
To quantify this difference, we fit both a simple power law of $K = K_{100}(r/100~\rm{kpc})^{\alpha}$ 
and the same power law plus a central entropy $K =K_0 +  K_{100}(r/100~\rm{kpc})^{\alpha}$ to the
entropy profiles, as was done for the active clusters in \citet{Donahue2005B}.  
Table~\ref{table:entropyfits}
gives the best fits.  We find that $\alpha \approx 0.6-0.8$ and $K_0 \approx 30-50 \,{\rm keV \, cm^2}$
in the radio-quiet clusters, in contrast to $\alpha \sim 1$ and $K_0 \approx 10 \,{\rm keV \, cm^2}$
for the active clusters.  Figure~\ref{figure:EntPower} shows central entropy values
plotted as a function of 20~cm radio power, from the NVSS \citep{NVSS}. Abell 2244 has a weak, off-center 
radio source that may not be associated with the cluster, plotted as an upper limit.

\begin{deluxetable}{lccccc}
\tablecaption{Entropy Profile Fit Results\label{table:entropyfits}}
\tablehead{
\colhead{Cluster} & \colhead{$K_0$} & \colhead{$K_{100}$} & \colhead{$\alpha$} & 
\colhead{Total $\chi^2$ }&\colhead{ N } \\
\colhead{}              & \colhead{KeV cm$^2$} & \colhead{KeV cm$^2$} & \colhead{}   & \colhead{}                          & \colhead{(d.o.f.)}  }
\startdata
Abell 1650 &  $27\pm5$ & $150\pm7$ & $0.80\pm0.07$ & 12 & 47 \\ 
                     &  $=0.00$    &  $177$         & $0.56\pm0.02$ &  28  & 48 \\
Abell 2244 &  $48\pm5$ & $ 102\pm8$ & $0.97\pm0.08$ &  7 & 31  \\ 
                     & $=0.00$     &  $162\pm3$         & $0.54\pm0.02$ & 42  & 32 \\ 
Active Sample & $8\pm4$ & $150\pm50$  & $1.2\pm0.2$ &   & \\
                           & $=0.00$\tablenotemark{*}  & $144\pm24$  & $0.96\pm0.15$ & & \\ 
\enddata
\tablenotetext{*}{The fits set to $0.00$ entropy in the cores for the sample in Donahue
et al. (2005B) were quite poor, except for the case of Abell 2029.}
\end{deluxetable}

\section{Discussion}

The significance of elevated central entropy in Abell~1650 and Abell~2244 is that
a larger central entropy implies a longer central cooling time compared to
clusters in \cite{Donahue2005B} of similar temperature.  Assuming pure free-free
cooling, the cooling time for gas of temperature $T$ and entropy $K$ is 
\begin{equation}
  t_c \approx 10^8 \, {\rm yr} \left( \frac {K} {10 \, {\rm keV cm^2}} \right)^{3/2} 
  				\left( \frac {kT} {5 \, {\rm keV}} \right)^{-1} \; \; .
\end{equation}
Thus, these two clusters, which show no evidence for feedback, have a central cooling 
time $\sim 1 \,{\rm Gyr}$, while those that do show evidence for feedback have a central
cooling time $\sim 0.1 \, {\rm Gyr}$.  According to the definition of \citet{Peres1998}, Abell~1650
and Abell~2244  were properly classified as cooling-flow clusters because $t_c < 5$~Gyr.  
However, one does not expect to see significant cooling and condensation of gas 
in these clusters for at least another $\sim 5 \times 10^8 \,{\rm yr}$.  In other words, evidence
for feedback is seen in those clusters that can trigger it on a $\sim 10^8 \, {\rm yr}$ timescale
and not in clusters in which gas is not currently expected to be condensing.  Here we discuss the
implications of this finding.

The most straightforward interpretation of the cooling-time dichotomy between active
and radio-quiet clusters is that radiative cooling in cluster cores triggers AGN feedback
when the central gas begins to condense.   \citet{Donahue2005B} find that all nine of 
their active clusters have very similar core entropy profiles, suggesting that this set 
of clusters has settled into a quasi-steady configuration that is episodically heated by 
AGN outbursts on a $\sim 10^8$~year timescale.   \citet{VoitDonahue2005} show that 
outflows of $\sim 10^{45} \, {\rm erg \, s^{-1}}$ naturally maintain the 
observed characteristics of the entropy profiles in these clusters.

If that is the correct interpretation, then it is possible that Abell~1650 and Abell~2244 have
unusually long cooling times because they each experienced unusually strong AGN outbursts
$\gtrsim 1$~Gyr in the past.  Raising the central entropy to the observed $\sim 30-50 \, 
{\rm keV cm^2}$ levels would require an AGN outflow $\sim 10^{46} \, {\rm erg \, s^{-1}}$ \citep{VoitDonahue2005}.  Such outbursts are rare but not unprecedented.  
\citet{McNamara2005} have recently observed an outburst of this magnitude in
MS0735+7421, which now has a central entropy $\sim 30 \, {\rm keV \, cm^2}$.
The long cooling time following such an outburst would account for why we do 
not see any sign of X-ray cavities in these two clusters.

It is also possible that the central gas in Abell~1650 and Abell~2244 has never cooled
to the point at which it can trigger a strong AGN outburst.  That could happen, for example, if
frequent merger shocks have been able to support the core entropy at the $\sim 50 \, 
{\rm keV \, cm^2}$ level for several Gyr, or if electron thermal conduction can resupply the
thermal energy radiated by the central gas.  One can evaluate the efficacy of thermal
conduction by comparing the size of a radiatively cooling system to the Field length
\begin{equation}
	\lambda_{\rm F} = \left( \frac {\kappa T} {n_e^2 \Lambda} \right)^{1/2}
	                              \approx 4 \, {\rm kpc} \, \left( \frac {K} {10 \,{\rm keV \, cm^2}} \right)^{3/2} f_c^{1/2} 
	                              \; \; ,
\end{equation}
where $\Lambda$ is the usual cooling function and $\kappa = 6 \times 10^{-7} \, f_c T^{5/2} 
\, {\rm erg \, s^{-1} \, cm^{-1} \, K^{-7/2}}$ is the Spitzer conduction coefficient with suppression 
factor $f_c$.  The approximation assumes free-free cooling ($\Lambda \propto T^{1/2}$), which conveniently
makes $\lambda_{\rm F}$ a function of entropy alone.  At radii $\sim 100 \, {\rm kpc}$, we find that
$\lambda_{\rm F} \sim r$ in all the cooling-flow systems we have studied, implying that conduction
can plausibly balance cooling there, as long as $f_c \sim 1$.  At radii $\sim 10 \, {\rm kpc}$ in systems
with signs of feedback, we find $\lambda_{\rm F} < r$ even for $f_c = 1$, implying that conduction
cannot balance cooling at small radii, in agreement with the findings of \citet{2004MNRAS.347.1130V}.
At those same small radii in the two systems without signs of feedback, we find $\lambda_{\rm F}
\approx r$ for $f_c \approx 1$, suggesting that these systems are potentially stabilized by
thermal conduction, which would account for their modest temperature gradients.

One speculation that emerges from this brief analysis of thermal conduction is that there
is a critical entropy profile $K(r) \approx 10 \, {\rm keV \, cm^2} \, f_c^{-1/3} (r/4 \, {\rm kpc})^{2/3}$
dividing conductively stabilized systems from those that require feedback.  Clusters with 
central entropy profiles below this line will continue to cool until some other heat source
intervenes, while conduction stabilizes those clusters above the line.  One would then expect
the cluster population to bifurcate into systems with strong central temperature gradients 
and feedback and those without either.   Furthermore, a very powerful AGN outburst
could induce a transition from a feedback-stabilized state to a conductively-stabilized state by 
raising the central entropy level to $\gtrsim 30 \, {\rm keV \, cm^2}$.

Another potential heat source that has been suggested as a solution to the cooling-flow problem 
is annihilation of dark matter particles such as neutralinos \citep{QinWu01,Totani2004}. In the 
model of \citet{Totani2004}, the annihilation rate peaks in the center because of a spike in the 
density profile owing to the central black hole.  The steady heating rate in this model is not linked 
as directly to baryon cooling as the AGN feedback model suggested here, but it is an interesting 
alternative mechanism  that could be explored further.

\section{Conclusions}
  In order to test whether AGN heating compensates for radiative cooling in the
  cores of clusters of galaxies, we have
used {\em Chandra} to observe a small sample consisting of two clusters with central cooling
times $< H_0^{-1}$,  
yet no evidence for prominent AGN activity:  Abell 1650 and Abell 2244. The X-ray properties of the 
cores of these clusters indeed appear systematically different from cores with
more prominent radio emission. While the central cooling times are shorter 
than a Hubble time and they have strong metallicity gradients,  
they do not have significant central temperature gradients,
and their central entropy levels are markedly higher than in clusters with stronger radio
emission, corresponding to central cooling times of a billion years.  Also, there is no
evidence in the X-ray surface brightness maps for fossil X-ray cavities produced by
a relatively recent episode of AGN heating.  In contrast to the central cores of the clusters 
with stronger radio emission, these cores may be stabilized by conduction if it is
operating at close the Spitzer rate. We suggest that a tremendous AGN outburst, such 
as that shocking the ICM in MS0735+74 \citep{McNamara2005} may have elevated the central
entropy of these clusters some $10^9$ years ago. Whether or not conduction is operative
in stabilizing these clusters cannot be determined, but it is energetically feasible. 
Further theoretical development and a larger study is required to test whether the
timescales are consistent with entropy profiles of a larger population. 
The fact that these clusters with no evident feedback have
higher central entropy than clusters with obvious feedback suggests that 
rare but influential AGN outbursts can dramatically change the original distribution of
entropy in clusters of galaxies. Alternatively, the intracluster gas of these clusters 
may have started out with higher initial 
entropy than the ICM in  the active clusters, and it has not cooled to the point of sparking strong
AGN feedback.

\acknowledgements
Support for this work was provided by the National Aeronautics and Space Administration through Chandra Award Numbers SAO GO3-4159X and AR3-4017A issued by the Chandra X-ray Observatory Center, which is operated by the Smithsonian Astrophysical Observatory for and on behalf of the National Aeronautics Space Administration under contract NAS8-03060.
This research has made use of the NASA/IPAC Extragalactic Database (NED) which is operated by the Jet Propulsion Laboratory, California Institute of Technology, under contract with the National Aeronautics and Space Administration.

\bibliography{coolingflows}

\end{document}